\documentclass[%
aps,prb,superscriptaddress,twocolumn,floatfix,10pt,longbibliography]{revtex4-2}
\usepackage[utf8]{inputenc}
\usepackage{amsmath,amssymb}
\usepackage{empheq}
\usepackage{float}
\usepackage{lipsum}
\usepackage{mathtools,cuted}
\usepackage{esvect}
\usepackage{multirow}
\usepackage{xcolor}
\usepackage{soul}
\usepackage[export]{adjustbox}
\usepackage{graphicx}
\usepackage{dcolumn}
\usepackage{bm}
\usepackage{physics}
\usepackage[mathlines]{lineno}
\usepackage{hyperref}
\usepackage[capitalise]{cleveref}
\usepackage{bbm}
\usepackage{orcidlink}

\begin{document}
	
\title{Anomalous Polarization in One-dimensional Aperiodic Insulators}

\author{A. Moustaj \orcidlink{0000-0002-9844-2987}}
    \affiliation{Institute of Theoretical Physics, Utrecht University, Princetonplein 5,  3584CC Utrecht, The Netherlands.}
\author{J.P.J. Krebbekx \orcidlink{0009-0009-3500-1656}}
    \affiliation{Institute of Theoretical Physics, Utrecht University, Princetonplein 5,  3584CC Utrecht, The Netherlands.}
    \author{C. Morais Smith \orcidlink{0000-0002-4190-3893}}
    \affiliation{Institute of Theoretical Physics, Utrecht University, Princetonplein 5,  3584CC Utrecht, The Netherlands.}

	\date{\today}
	
\begin{abstract}

Multilevel charge pumping is a feature that was recently observed in quasiperiodic systems. In this work, we show that it is more generic and appears in different aperiodic systems. Additionally, we show that for aperiodic systems admitting arbitrarily long palindromic factors, the charge pumping protocol connects two topologically distinct insulating phases. This confirms the existence of topological phases in aperiodic systems whenever their finite-size realizations admit inversion symmetry. These phases are characterized by an anomalous edge response resulting from the bulk-boundary correspondence. We show that these signatures are all present in various chains, each representing a different class of structural aperiodicity: the Fibonacci quasicrystal, the Tribonacci quasicrystal, and the Thue-Morse chain. More specifically, we calculate three quantities: the Berry phase of the periodic approximation of the finite-size systems, the polarization response to an infinitesimal static and constant electric field in systems with open boundary conditions, and the degeneracy of the entanglement spectrum. We find that all of them provide signatures of a topologically nontrivial phase.
\end{abstract}
	
	\maketitle{}
	
	
\section{Introduction}\label{sec:Intro}
Electronic band theory is a cornerstone of condensed-matter physics, usually grounded on the crystalline structures found in solid-state materials. The latter allows one to use crystal cells to construct Bloch Hamiltonians, which are much simpler to deal with than their real-space counterparts. By using these Bloch band structures, a well-established theoretical framework for topological insulators (TI's) and superconductors has been developed during the last decades \cite{Schnyder2008ClassificationDimensions,Ryu2010TopologicalHierarchy,Ludwig2016TopologicalBeyond}. However, topologically nontrivial insulating states are not restricted to periodic systems, as they have also been found in amorphous materials \cite{Agarwala2017TopologicalSystems,Grushin2022TopologicalMatter}. 

One of the main properties of TI's is the bulk-boundary correspondence, which implies the existence of robust in-gap edge modes in one dimension (1D) and \textit{gapless} boundary states in two and three dimensions (2D and 3D). The physical response of such systems can be probed by the application of external perturbations, after which one finds either conducting channels at the boundary of an otherwise insulating bulk in 2D and 3D, or an anomalous polarization response and fractional corner charges in 1D \cite{Su1979SolitonsPolyacetylene,Cheng2023TransportModels,Zak1989BerrysSolids,Aihara2020AnomalousPhase}. Another class of 1D systems for which quantized boundary responses are possible are superconductors. A prime example where this happens is the Kitaev chain \cite{Kitaev2001UnpairedWires}, which hosts Majorana edge states.
\\

In the case of periodic crystals, the nontrivial topology of 2D and 3D TI's is generally attributed to the nonexistence of a complete set of exponentially localized Wannier functions, also known as an \textit{obstruction to Wannierization} \cite{Brouder2007ExponentialInsulators}. As a consequence, a topologically nontrivial state cannot be adiabatically connected to an atomic limit in which electrons occupy maximally localized Wannier states. In 1D, however, it is known that localized Wannier bases can always be constructed, such that there cannot be any topological obstruction \cite{Kohn1959AnalyticFunctions}. Nevertheless, topologically nontrivial insulators, with anomalous polarization and fractional boundary charges, in fact exists. These phases are known to be equivalent to \textit{obstructed atomic limits}, which are not adiabatically connected with \textit{trivial atomic limits} \cite{Bradlyn2017TopologicalChemistry}. In order for this to happen, crystalline symmetries must be imposed on the system. The only possible such symmetry in 1D is inversion, forcing the polarization density of the system to be quantized at 0 or 1/2 (in units where $e=1$) \cite{Zak1989BerrysSolids,Aihara2020AnomalousPhase}. In turn, it is possible to relate the bulk topological invariant of a chiral-symmetric (or equivalently sublattice-symmetric) 1D system to its quantized polarization \cite{Chiu2016ClassificationSymmetries}, and therefore understand the physical implications of the nontrivial topology of a 1D insulator. Namely, there exists an anomalous boundary response in the obstructed atomic phase in the form of a quantized polarization resulting from the pile up of fractionalized charges at the edges. The paradigmatic example of such a phase is realized in the Su-Schrieffer-Heeger (SSH) model \cite{Su1979SolitonsPolyacetylene}. \\

Given that most toy models exhibiting such phases are typically periodic crystals, one wonders whether aperiodic 1D systems can also host obstructed atomic limits. Intense efforts have been made in the last decade to understand whether nontrivial topological phases could also exist in 1D quasicrystalline systems \cite{Kraus2012TopologicalModel,Kraus2012TopologicalQuasicrystals,Madsen2013TopologicalStructures,Levy2016TopologicalNumbers}. It was shown that the gapped spectra of quasicrystalline Hamiltonians carry nontrivial signatures of topology. For example, a typical quasicrystal carries gap labels which have a topological origin and can be understood from K-theory \cite{Bellissard1992GapOperators}. These topological indices have also been linked with the pumping of boundary states \cite{Kraus2012TopologicalQuasicrystals,Verbin2015TopologicalQuasicrystal}.
However, the question remains whether these observations constitute genuine signatures of anomalous boundary physics, related to bulk properties by means of a bulk-boundary correspondence. In other words, do the topological labels observed in noncrystalline systems in 1D imply the existence of an electronically insulating topological phase, which is adiabatically connected to some type of obstructed atomic limit, and to which anomalous boundary physics can they be linked to? To the best of our knowledge, this question has so far not been fully addressed, and we show in this work that it is indeed possible to find such phases for finite-size realizations of aperiodic systems. In order to reach this conclusion, we will use typical probes for anomalous edge responses, such as the Berry phases of crystalline approximants, the anomalous polarization responses to external static electric fields and entanglement spectrum (ES) degeneracies. The systems that we will consider are the Fibonacci chain, the Tribonacci chain, and the Thue-Morse chain. 
By performing the adiabatic charge pumping protocol proposed in Ref.~\cite{Yoshii2021TopologicalIndex} for the Fibonacci chain, we show that multilevel pumping also works for at least two other aperiodic modulations: the Tribonacci sequence and the Thue-Morse sequence. We then identify the points in time where the Berry phase of these approximants is exactly $0$ or $\pi$. These turn out to be the points where the boundary states of open systems are degenerate. For the Thue-Morse chain, a similar situation happens as for the SSH model. The $\phi=\pi$ point is, in fact, an inversion-symmetric realization of the system, which admits an anomalous polarization response of $P=0.5$ due to the degeneracy of the edge modes. For the Fibonacci and Tribonacci chains, we use their palindromic factors to study their inversion-symmetric realizations and draw the same conclusions as before. 
\\

This work is structured as follows. In \cref{Sec: TIs and signatures}, we briefly review the classification scheme of topological insulators and the associated physical observables that result from the bulk-boundary correspondence in 1D. This is followed in \cref{Sec: Aperiodic Systems} by a short overview of aperiodic systems.
In \cref{Sec: TCPAS}, we present the charge-pumping protocol and the results obtained for the various aperiodic systems considered in this work. 
In \cref{Sec: Topo Signatures Aperiodic}, the different physical signatures of 1D TI's in three different classes of aperiodic systems are discussed. We show that all the signatures that one should expect to see are present in these systems. Finally, in \cref{Sec: Conclusion}, we conclude with a summary and outlook.

\section{Signatures of 1D Topological Insulators}\label{Sec: TIs and signatures} 
Here, we briefly review the concept of a topological insulator. Their theoretical understanding is based on a classification in terms of anti-unitary symmetries and dimensionality. This classification is equivalent to that of random matrices, and is known as the Altland-Zirnbauer classification, or the ten-fold way \cite{Altland1997NonstandardStructures}.
 
The most general statement that can be made concerning symmetry-protected topological states of matter is that they are ground states of a many-body system that are adiabatically distinct from a trivial product of single-particle states, also called a \emph{trivial atomic limit}.  
By using K-theory, it has been shown that the generalized homotopy group of a Bloch Hamiltonian taking values in the $d$-dimensional Brillouin zone (BZ) $T^d$ and belonging to a certain classifying space $R_q$ can be written as \cite{Kitaev2009PeriodicSuperconductors}
\begin{equation*}
    \pi(T^d,R_q)=\pi_0(R_{q-d})\bigoplus_{s=0}^{d-1}\begin{pmatrix}d \\ s\end{pmatrix}\pi_0(R_{q-s}).
\end{equation*}
The classifying spaces $R_q$ are the sets of real symmetric matrices under the constraint of anti-unitary symmetries and with eigenvalues $\pm1$. They are labeled by the integer $q\mod8$ and correspond to the eight real Cartan classes. A similar equation can be posed for the complex classifying spaces $C_q$, corresponding to the set of Hermitian matrices labeled by the integer $q\mod2$, and associated with the two complex Cartan classes. The first part of this equation, namely $\pi_0(R_{q-d})$ is what appears in the ten-fold classification. The second part is attributed to unitary symmetries, which allow point-group symmetries to protect topologically nontrivial states as well. A consequence of this is the existence of topological crystalline insulators \cite{Fu2011TopologicalInsulators}, which are nontrivial topological states protected by crystalline symmetries of the lattice.
Now, let us review some of the tools developed to characterize topological states of matter in 1D. 

We consider a two-band, 1D periodic crystal, for which the insulating ground state consists of a filled valence band, separated from the conduction band by a gap. The paradigmatic model for such a system is the SSH chain \cite{Su1979SolitonsPolyacetylene}. It will be used as a benchmark in \cref{Sec: Topo Signatures Aperiodic} to compare the results on aperiodic systems. The Hamiltonian is given by 
\begin{equation} \label{Eq: SSH chain}
    H_{\text{SSH}}=-g\sum_{j=1}^{2N}\left[1-(-1)^j\delta\right]c_j^\dagger c_{j+1}+\text{h.c.},
\end{equation}
where $g$ is the hopping amplitude, $\delta$ is the dimerization parameter, and $N$ the amount of unit cells. We can define the Berry phase of the valence band eigenstate $\ket{\psi_-(k)}$ with respect to the BZ as 
\begin{equation*}
    \phi=i\oint_{\text{BZ}}dk\bra{\psi_-(k)}\partial_k\ket{\psi_-(k)}.
\end{equation*}
It turns out that under the constraint of inversion symmetry in 1D, the Berry phase (also called Zak phase in this case) must be quantized to either 0 or $\pi$ \cite{Zak1989BerrysSolids}. 
Since the Hamiltonian also exhibits chiral symmetry $\mathcal{S}H\mathcal{S}^{-1}=-H$, with $\mathcal{S}$ denoting the chiral symmetry operator, the boundary modes are forced to be at $E=0$, which provides further robustness to the topological phase. This spectral symmetry allows one to define the winding number \cite{Chiu2016ClassificationSymmetries}
\begin{equation*}
    \nu = \int\frac{dk}{4\pi i}\Tr\left[\mathcal{S}\partial_k\log H(k)\right],
\end{equation*}
which can be shown to be related to the quantized Berry phase of an inversion-symmetric system through $\phi=\pi\nu\mod{2\pi}$. For $\delta<0$, the SSH chain is in the topological phase and has two boundary zero modes. The winding number can be calculated exactly and is $\nu=0$ ($\nu=1$) for $\delta>0$ ($\delta<0$).

The quantization of the winding number has an interesting physical meaning, which eventually takes the form of a bulk-boundary correspondence. The reason is that the Berry phase measures the average position of electrons in the ground state within a unit cell \cite{King-Smith1993TheorySolids}. In the case of an inversion-symmetric insulator, $\phi=0$ corresponds to a charge distribution around the center of a unit cell, while $\phi=\pi$ corresponds to a charge distribution at its boundaries. The eigenvalues of the polarization density operator (in units of $e=1$) are related to the Berry phase through the following equation \cite{King-Smith1993TheorySolids} 
\begin{equation}
    P=\frac{\phi}{2\pi }\mod{1}=\left\{\begin{alignedat}{5}
    & 0, \ \ \ \text{if } \phi=0, \\
               & \frac{1}{2}, \ \ \ \text{if } \phi=\pi.
\end{alignedat}\right.
\end{equation}
This means that if one takes a finite system composed of an integer amount of unit cells to keep the inversion symmetry intact, one finds half charges located at the boundaries of the system. The response to an infinitesimal external electric field should reflect this property. If a system does not preserve inversion symmetry, then the polarization density can take any value. This means that it is only possible to obtain quantized Berry phases and, therefore, a quantized boundary response when inversion symmetry is present in 1D. 

We can further understand these topological properties in terms of adiabatic deformations from trivial atomic limits to obstructed atomic limits. They are the only two different elementary band representations for systems possessing inversion symmetry \cite{Bradlyn2017TopologicalChemistry}. The half charges that result from such a situation are also intimately related to the concept of a filling anomaly, in which Wannier centers do not coincide with atomic sites \cite{Schindler2020Higher-OrderMatter}. Under open boundary conditions, this results in degenerate boundary states at the Fermi energy.
Finally, this last point allows another useful formulation of the topological phase through the ES. A robust signature of this filling anomaly is a degenerate ES of the insulator's many-body ground state \cite{Fidkowski2010EntanglementSuperconductors}.

Most of the theory discussed in this section relies on translational invariance and a crystalline structure for theoretical analysis. However, our focus is on understanding how these features manifest in aperiodic systems. To facilitate this exploration, we will first provide a brief overview of these systems in the following section.

\section{Aperiodic Systems}\label{Sec: Aperiodic Systems}

Aperiodic sequences have attracted considerable interest since their first applications in physical systems. For example, by studying spin chains under the influence of aperiodic modulations, it was understood that depending on the fluctuations of the aperiodic sequence, it is possible to access different universality classes of quantum phase transitions \cite{Harris1974EffectModels,Luck1993CriticalField,Hermisson2000AperiodicResults}. Later on, the properties of single-particle Hamiltonians on 1D quasicrystals were investigated using various techniques, such as renormalization schemes and trace maps. Amongst many interesting results, it was discovered that the spectrum is singular continuous and the eigenstates are neither localized nor delocalized, but are \textit{critical} and exhibit multifractal behavior \cite{Kohmoto1984CantorMap,Bellissard1989SpectralQuasi-Crystals,Sire1989SpectrumChain,Sire1990ExcitationQuasicrystals,Niu1991SpectralStructures}. More recently, theoretical and experimental works have shown that: i) it is possible to observe topological charge pumping in quasiperiodic chains \cite{Kraus2012TopologicalQuasicrystals,Verbin2013ObservationQuasicrystals,Verbin2015TopologicalQuasicrystal}; ii) that the critical eigenstates are robust against local impurities \cite{Moustaj2021EffectsQuasicrystal}; iii) That the critical states emerge from a cascade of localization-delocalization transitions \cite{Goblot2020EmergenceChains}; iv) coupling of quasicrystalline chains may result in the simultaneous existence of critical and extended eigenstates \cite{Moustaj2023SpectralChains}. It was also shown that it is possible to control edge states by manipulating local structures in various aperiodic systems due to the presence of local symmetries \cite{Rontgen2019LocalChains}. 
Aperiodic modulations may have other interesting implications on physical systems. We refer the reader for a comprehensive review to Refs.~\cite{Macia2006TheTechnology,Jagannathan2021TheMultifractality}

Contrary to their periodic counterparts, these systems do not possess translation symmetry and are sometimes referred to as systems with deterministic disorder. They show some type of long-range order even though they are not periodic \cite{Macia2006TheTechnology}. In 1D, this \textit{aperiodic} order is usually encoded in sequences of symbols that can be generated using an inflation rule. The mathematical study of such structures is called symbolic dynamics \cite{Brin2002SymbolicDynamics}. That is the study of the dynamical system formed by a set of symbols subjected to the repeated application of a function that maps the set of symbols to the set of combinations of symbols. 

We consider finite-size words $W\in \mathcal{V}$, where $\mathcal{V}$ denotes the set of finite words that can be generated from an alphabet $\mathcal{A}=\{a_0,\ a_1,\ \cdots, a_m\}$ of inequivalent symbols. These words can be constructed by repeatedly applying a substitution rule $\sigma: \mathcal{A}\to \mathcal{V}$. This rule imposes recurrence relations on words, which makes it easy to generate them. 
Suppose one starts with the ``seed" letter $a_0$. We call the word \textit{uniquely} generated by applying the rule $\sigma$ to the seed letter $n$ times the $n^{\text{th}}$ approximant of the aperiodic sequence. Below, we give a few examples of aperiodic sequences that we use throughout this work.

\paragraph{Fibonacci sequence.} The Fibonacci words can be generated from the binary alphabet $\mathcal{A}=\{A,B\}$ by applying the following recursion relation, 
\begin{align*}
    W_n&=W_{n-1}W_{n-2}, \ \text{for} \ n>1, \\
    W_0&=A, \ W_1=AB,
\end{align*}
In the above equation, the product of words means that they are concatenated. This is an example of a 1D quasicrystal, as it can be obtained via a cut-and-project scheme from a regular 2D square lattice \cite{Baake2013AperiodicOrder}. 

\paragraph{Tribonacci sequence.} The Tribonacci word is an extension of the Fibonacci word, and it is also a quasicrystal. However, its cut-and-project scheme results from a 3D cubic lattice instead \cite{Krebbekx2023MultifractalChains}. The alphabet generating the word is composed of three letters $\mathcal{A}=\{A,B,C\}$, and the recursive scheme to generate the word is 
\begin{align*}
    W_n&=W_{n-1}W_{n-2}W_{n-3}, \ \text{for} \ n>2, \\
    W_0&=A, \ W_1=AB, \ W_2=ABAC.
\end{align*}

\paragraph{Thue-Morse sequence.} The two previous words are examples for which the Pisot substitution conjecture holds \cite{Akiyama2015OnConjecture,Adiceam2016OpenQuasicrystals}. The characteristic matrix of the substitution dynamics has a polynomial of degree equal to its dimension. This makes their diffraction spectrum pure-point. For this reason, they can be called quasicrystals \cite{Baake2013AperiodicOrder}. The Thue-Morse chain is not generated by a Pisot substitution and is an example of an aperiodic chain that is not a quasicrystal. It has a singular-continuous diffraction spectrum \cite{Baake2019ScalingMeasure}. The sequence is generated by repeatedly applying the substitution rule $\sigma(A)=AB$ and $\sigma(B)=BA$ to the binary alphabet. Alternatively, it can be generated by the following recurrence relation 
\begin{align*}
    W_n&=W_{n-1}\overline{W_{n-1}}, \ n>0, \\
    W_0&=A,
\end{align*}
where $\overline{W_n}$ is the bit-wise negated word: $A\to B$ and $B\to A$. That is, the $n^{\text{th}}$ generation word is obtained by concatenating the $n-1$ generation word with its bit-wise negated version. 

\paragraph{Rudin-Shapiro sequence.}
Finally, the last example of an aperiodic chain that we investigate is the Rudin-Shapiro chain \cite{Dulea1992Trace-mapModel}. This system differs from the previous three because it features an absolutely continuous diffraction spectrum. In order to generate the binary Rudin-Shapiro sequence, we use the following two-step procedure: We first impose the substitution rule on the four-letter alphabet $\mathcal{A}=\{A,B,C,D\}$:
\begin{equation*}
    \sigma:\left\{\begin{alignedat}{5}
    & A\mapsto AB,  \\
    & B\mapsto AC, \\
    & C\mapsto DB, \\
    & D\mapsto DC.
\end{alignedat}\right.
\end{equation*}
This is then followed by the second step, which identifies $A,B\to A$, and $C,D\to B$. This way, starting with the seed $A$, we obtain the Rudin-Shapiro binary word $AAABAABAAAABBBAB\cdots$. 

In the following section, we will explore how topological charge pumping is achieved in these aperiodic systems. 

\section{Topological Charge Pumping of Aperiodic Systems}\label{Sec: TCPAS}

The idea of topological charge pumping dates back to an argument made by Laughlin to explain the robustness of the quantized conductance in a two-dimensional quantum Hall system \cite{Laughlin1981QuantizedDimensions}. Soon thereafter, Thouless showed, using similar arguments, that particle transport in a 1D system subjected to an adiabatic evolution will obey the same quantization condition \cite{Thouless1983QuantizationTransport}. Because charge pumping is analogous to the Quantum Hall phase in 2D, where a quantized amount of charge is pumped on the boundary of a cylinder upon the insertion of one flux quantum \cite{Laughlin1981QuantizedDimensions,Thouless1982QuantizedPotential}, this provides an intuitive understanding of the dimensional extension performed in Ref.~\cite{Kraus2012TopologicalQuasicrystals}.
This extension, achieved with the 2D parent Hamiltonian representation, reveals that the topological classification enjoyed by the 1D quasicrystalline insulator is equivalent to the 2D class $A$ of the ten-fold way \cite{Altland1997NonstandardStructures}.
Recent work has also shown quantized topological charge transport in a periodically modulated (in time)  1D Fibonacci quasicrystal \cite{Yoshii2021TopologicalIndex}. A Rice-Mele \cite{Rice1982ElementaryPolymer} pump with a Fibonacci modulation of the potentials displays the interesting possibility of multilevel pumping, a feature that is not present in equivalent crystalline systems. 
When the modulation is quasiperiodic, the number of pumped charges is instead characterized by a Bott index \cite{Loring2010DisorderedC-algebras,Hastings2010AlmostEffect}, which, in the thermodynamic limit, is equivalent to the Chern number \cite{Thouless1982QuantizedPotential} obtained in the periodic modulation case.

The symmetry class $A$ is the most robust class, as it is devoid of any symmetry. As long as the perturbations are applied adiabatically, the system is guaranteed to stay in the same topological phase. This is the reason why quasi-crystallinity is completely irrelevant to the topological classification of a 2D time-reversal-symmetry-breaking insulator. This will also be shown in the next sections when we perform adiabatic charge pumping to different classes of aperiodic modulations. 

We will next briefly describe how quantized charge pumping occurs in a 1D periodic crystal and how this quantization can be ascribed to the Chern theorem. 

\subsection{Charge transport as a polarization current}
Starting from a one-dimensional crystalline insulator, an adiabatic evolution of the system driven by a parameter $\lambda$ induces a change in the polarization density
\begin{equation*}
    \Delta P = \int_{\lambda_i}^{\lambda_f} d\lambda\partial_\lambda P.
\end{equation*}
The modern theory of polarization \cite{King-Smith1993TheorySolids}
dictates that
\begin{equation*}
    \partial_\lambda P= \frac{1}{2\pi}\sum_{n}^{\text{occ}}\int_0^{2\pi}dk \ \text{Im}\bra{\partial_\lambda u_{n}}\ket{\partial_ku_{n}}
\end{equation*}
where $\ket{u_n}$ is a Bloch state corresponding to the $n^{\text{th}}$ band of the system. If the parameter $\lambda$ is periodic, i.e. the evolution is cyclic, one can identify $\text{Im}\bra{\partial_\lambda u_{n}}\ket{\partial_ku_{n}}=F^{(n)}(\lambda,k)$ as a Berry curvature, which means that the total change in polarization over one cyclic change of $\lambda$ yields a total integral of the Berry curvature over a torus,
\begin{equation}\label{Eq: Total Change Polarization Chern}
\begin{split}
    \Delta P&=\frac{1}{2\pi}\sum_{n}^{\text{occ}}\int_{0}^{2\pi} dk\oint d\lambda F^{(n)}(\lambda,k) \\
    &= \sum_{n}^{\text{occ}}C_n\equiv C\in \mathbb{Z},
\end{split}
\end{equation}
where $C_n$ denotes the Chern number of the $n^{th}$ band, given by \cite{Thouless1982QuantizedPotential}
\begin{equation}\label{Eq: Chern number}
    C_n=\int\frac{dk_xdk_y}{2\pi}F^{(n)}_{xy}(\mathbf{k}),
    \end{equation}
where $F^{(n)}_{xy}(\mathbf{k})=\partial_x\mathcal{A}^{(n)}_y-\partial_y\mathcal{A}^{(n)}_x$ is the Berry curvature, $\mathcal{A}^{(n)}_\alpha(\mathbf{k})=\bra{u_n(\mathbf{k})}\partial_\alpha\ket{u_n(\mathbf{k})}$ is the Berry connection corresponding to the $n^{\text{th}}$ band. Note that we relabeled $k$ as $k_x$ and $\lambda$ as $k_y$ to be consistent with the literature on 2D topological band insulators. 

In the modern theory of polarization, this change can also be understood as a change in the Wannier center positions through the Berry phase formulation of the Wannier center. Therefore, \cref{Eq: Total Change Polarization Chern} predicts that there is an integer amount of charges crossing the unit cell of the one-dimensional crystal. This means that, as a consequence of the Chern theorem, the total change in polarization is quantized to an integer number.

An intuitive example of a system that will also be discussed later is the Rice-Mele charge pump \cite{Rice1982ElementaryPolymer,Asboth2016AInsulators}. This model connects the trivial phase of the SSH model to its topological phase and back to its trivial phase in an adiabatic cycle by breaking chiral symmetry. 
The Rice-Mele Hamiltonian is given by 
\begin{equation}\label{eq: Rice-Mele real-space}
\begin{split}
    \begin{multlined} H(t) = \sum_{j=1}^N\left[\Delta-(-1)^j\delta(t)\right]c^\dagger_jc_{j+1}
       \\ - \sum_{j=1}^N(-1)^jh(t)c^\dagger_jc_{j} + \text{h.c.},
    \end{multlined}
\end{split}
\end{equation}
where the first part is similar to the SSH model in \cref{Eq: SSH chain}, but with an independent hopping parameter $\Delta$ and a time-dependent dimerization $\delta(t)$. The second term adds a time-dependent staggered on-site potential $h(t)$, such that the bulk gap stays open throughout the whole period. The time-modulated functions are given by 
\begin{align*}
        \delta(t)&\equiv \delta_0 \cos(2\pi t/T), \\
        h(t)&\equiv h_0 \sin(2\pi t/T).
\end{align*}
Here, $\delta_0$ and $h_0$ are constant amplitudes, and $T$ is the modulation period. The Bloch Hamiltonian for this system is simply given by 
\begin{equation*}
    H(k,t)=\begin{pmatrix}
        h(t) & v_++ v_-e^{-ik} \\ v_+ + v_-e^{ik} & -h(t)
    \end{pmatrix},
\end{equation*}
where $v_\pm  = \Delta \pm \delta(t)$. Then, it becomes clear that at $t=0,T$, the system is in the trivial phase of the SSH model, with $h(0)=0$ and $v_+>v_-$. On the other hand, at $t=T/2$, the system is in the topological phase, with $v_+<v_-$.

\subsection{Bott-Index formulation of quantized charge pumping}
In Ref.~\cite{Yoshii2021TopologicalIndex}, the Bott index \cite{Loring2010DisorderedC-algebras,Hastings2010AlmostEffect} was used to demonstrate quantized charge pumping for a generalized Fibonacci chain, whose on-site and hopping parameters followed the quasiperiodic modulation. We will now summarize how the Bott index is defined for charge pumping and show that it leads to the same phase diagram as the Chern number.

\begin{figure}
    \centering
    \includegraphics{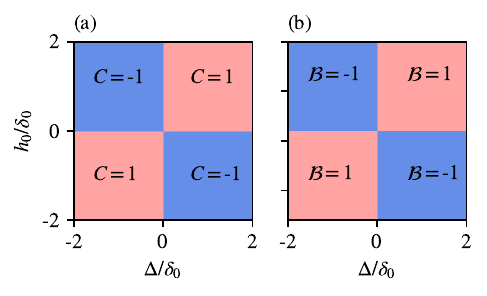}
    \caption{Equivalence of the topological phase diagram of the periodic Rice-Mele model, at half-filling, between (a) the Chern number formulation and (b) the Bott index formulation.}
    \label{fig: Chern vs Bott RM}
\end{figure}

In its most general form, the Bott index is a measure of the total phase picked up by a string of 2D position operators ($\hat{X}, \hat{Y})$, projected onto the insulating ground state $\ket{\psi}$, as they complete infinitesimal loops in real space. The contribution from all such loops over the whole system yields the Bott index. More concretely, let 
\begin{equation}\label{eq: UV operators Bott index}
    \begin{split}
        \hat{U}&=\hat{\mathcal{P}}\exp\left(\frac{2\pi i\hat{X}}{L_x}\right)\hat{\mathcal{P}}, \\
        \hat{V}&=\hat{\mathcal{P}}\exp\left(\frac{2\pi i\hat{Y}}{L_y}\right)\hat{\mathcal{P}},
    \end{split}
\end{equation}
where $L_x$ and $L_y$ are the dimensions of the system in the $x$ and $y$ directions, respectively.
The Bott index is then defined as 
\begin{equation}
    \mathcal{B}\equiv \frac{1}{2\pi}\Im\Tr\log\left(\hat{V}\hat{U}\hat{V}^\dagger\hat{U}^\dagger\right).
\end{equation}
It has been shown that the Bott index $\mathcal{B}$ is equal to the Chern number $C$ in the thermodynamic limit \cite{Loring2010DisorderedC-algebras}. However, even in finite-size systems, under periodic boundary conditions, it is a good indicator of a nontrivial topological character. Indeed, it works very well for nonperiodic 2D systems, such as disordered Chern insulators or amorphous materials \cite{Grushin2022TopologicalMatter,Loring2010DisorderedC-algebras,Agarwala2017TopologicalSystems}. In the case of adiabatic charge pumping, one deals with a periodic temporal parameter, which simplifies the formulation of the problem, as the Hamiltonian is in block diagonal form along the time axis and has instantaneous eigenstates $\ket{\psi(t)}$. It can be shown that the Bott index can be formulated in terms of a new set of operators $\Tilde{U},\Tilde{V}$, which are obtained from \cref{eq: UV operators Bott index} by understanding the action of the $\hat{Y}$ operator on momentum eigenstates. These operators take the form
\begin{equation*}
    \begin{split}
        [\Tilde{U}_t]_{mn}&=\bra{\psi_m(t)}\exp\left(\frac{2\pi i\hat{X}}{L_x}\right)\ket{\psi_n(t)}, \\
        [\Tilde{V}_{t,t+\Delta t}]_{mn}&=\bra{\psi_m(t)}\ket{\psi_n(t+\Delta t)},
    \end{split}
\end{equation*}
where $\Delta t$ denotes a discrete time step between adjacent times. Therefore, one can compute the Bott index as 
\begin{equation} \label{eq: Time Bott Index}
    \mathcal{B}=\frac{1}{2\pi}\sum_{t=0}^T \Im\Tr\log(\Tilde{V}_{t,t+\Delta t}\Tilde{U}_{t+\Delta t}\Tilde{V}^\dagger_{t,t+\Delta t}\Tilde{U}^\dagger_{t}).
\end{equation}
In \cref{fig: Chern vs Bott RM}, one can see the equivalence of the phase diagram computed from (a) the Chern number given by \cref{Eq: Chern number} and (b) the Bott index given by \cref{eq: Time Bott Index}.

\begin{figure*}[!hbt]
    \centering
    \includegraphics{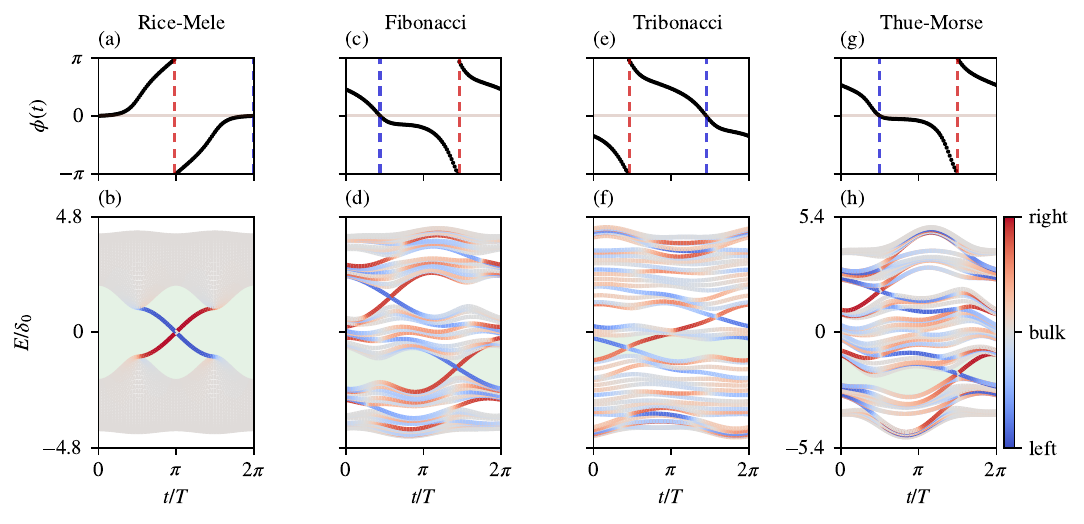}
    \caption{Top row: time evolution of the Berry phases of each model studied under a pumping cycle. Vertical dashed lines indicate points when $\phi=0$ (blue) or $\phi=\pi$ (red). Bottom row: behavior of the eigenvalues under adiabatic time evolution. The colors indicate the localization behavior of each mode, with grey denoting bulk modes and red (blue) indicating localization on the right (left) of the chain. The green shaded area indicates the bulk gap for which the Berry phase was calculated. (a,b) Periodic Rice-Mele, (c,d) Fibonacci Rice-Mele, (e,f) Tribonacci Rice-Mele, (g,h) Thue-Morse Rice-Mele. System sizes are (a,b) $N=100$, (c,d) $N=55$, (e,f) $N=44$, and (g,h) $N=64$.}
    \label{fig: Zak Phase and Gap Modes}
\end{figure*} 
\subsection{Aperiodic systems}

Recently, a system described by \cref{eq: Rice-Mele real-space} was investigated, but the periodic modulation of the dimerization and on-site potential was modified to a quasiperiodic one, following the Fibonacci sequence \cite{Yoshii2021TopologicalIndex}. It was shown that quantized topological charge pumping occurs in a time-periodically modulated 1D Fibonacci quasicrystal. In addition, multilevel pumping, i.e., pumping across multiple gaps simultaneously, was shown to be possible. This feature is not present in equivalent crystalline systems. This can be understood from a renormalization perspective, in which gaps of different generations can be mapped to each other due to the self-similar nature of the energy spectrum. In this work, we show that multilevel pumping occurs for at least two other aperiodic modulations: a Tribonacci sequence and a Thue-Morse sequence. Each one represents a different class of aperiodicity, with the Tribonacci chain being a quasicrystal obtained from projecting a 3D cubic lattice onto a line with an irrational slope and the Thue-Morse being an aperiodic sequence not forming a quasicrystal. Since topological charge pumping is independent of the specific realization of spatial symmetries, we will first examine aperiodic sequences generated by their standard rule. In \cref{Sec: Topo Signatures Aperiodic}, however, we will explore the palindromic versions of these sequences to study obstructed insulating states.
The Rice-Mele Hamiltonian in \cref{eq: Rice-Mele real-space} is modified to
\begin{align}\label{eq: Yoshi-Rice-Mele Hamiltonian}
    H(t)&=
        \sum_{j=1}^N\left(\left\{\left[\Delta-V_j\delta(t)\right]c^\dagger_jc_{j+1}+\text{h.c.}\right\} - V_jh(t)c^\dagger_jc_{j}\right),
\end{align}
where $V_j$ is the $j^{\text{th}}$ component of the aperiodic sequence of potentials distributed according to the aperiodic word $W_n$, with $N=|W_n|$ sites for periodic boundary conditions (PBC) and $N-1$ sites for open boundary conditions (OBC). 
We will consider finite-size words $W_n\in \mathcal{
V}$ generated by a substitution rule, where $\mathcal{
V}$ denotes the set of finite words that can be generated from an alphabet $\mathcal{A}$ (see \cref{Sec: Aperiodic Systems}). 

\begin{figure}[!hbt]
    \centering
    \includegraphics{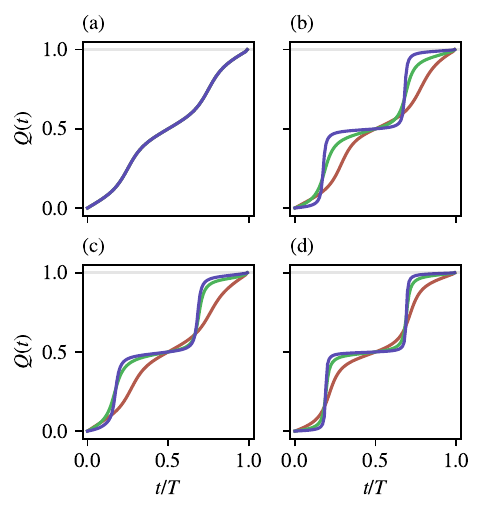}
    \caption{Cumulative charge pumped as a function of time. (a) The periodic modulation at half-filling. (b) The Fibonacci modulation at fillings $\tau$ (red), $\tau^3$ (green), and $\tau^6$ (blue), with $\tau = (\sqrt{5}-1)/2$ the inverse of the golden ratio. (c) The Tribonacci modulation at fillings $\beta$ (red), $\beta^3$ (green), and $\beta^4$ (blue), with $\beta^{-1} \approx 1.8393$ the real root of the cubic equation $x^3-x^2-x-1=0$, also called the Tribonacci constant. (d) The Thue-Morse modulation at fillings 1/3 (red), 1/10 (green), and 1/27 (blue). For the aperiodic models, the behavior tends to a step-like function for smaller fillings, which is a feature of multi-level charge pumping. System sizes are (a) $N=100$, (b) $N=55$, (c) $N=81$, and (d) $N=256$.}
    \label{Fig: Cumulative Charge All models}
\end{figure}

Now, we apply the first three modulations described in \cref{Sec: Aperiodic Systems} to the Hamiltonian \cref{eq: Yoshi-Rice-Mele Hamiltonian}, and numerically compute the time-dependent spectra. The results are shown in \cref{fig: Zak Phase and Gap Modes}. Additionally, we calculate the time-dependent Berry phases of the crystalline approximant in the insulating state with a band gap shown in green in \cref{fig: Zak Phase and Gap Modes}, which can possibly indicate a crystalline topological phase at the point where the in-gap edge modes cross and become degenerate. 
In Figs.~\ref{fig: Zak Phase and Gap Modes}(a) and \ref{fig: Zak Phase and Gap Modes}(b), the results for the periodic modulation are shown. In \cref{fig: Zak Phase and Gap Modes}(a), we see that at $t=T/2$, the Berry phase is equal to $\pi$, indicating the presence of inversion symmetry \cite{Zak1989BerrysSolids}. This result is corroborated by the crossing of the in-gap modes at $E=0$ in \cref{fig: Zak Phase and Gap Modes}(b). The color code of the eigenstates indicates whether they are bulk (grey)- or edge-localized (red and blue) modes. This confirms that at $t=T/2$, the chain is in the topological phase. These results are consistent with the fact that the Rice-Mele model realizes topological charge pumping by adiabatically connecting the topological and trivial phases of the SSH model.
In Figs.~\ref{fig: Zak Phase and Gap Modes}(c) and \ref{fig: Zak Phase and Gap Modes}(d), the same information is plotted for the Fibonacci modulation. In \cref{fig: Zak Phase and Gap Modes}(c), one can see that the edge modes now intersect at $t=T/4 $ and $t=T/4 $. However, contrary to the SSH model, the Fibonacci chain does not become inversion symmetric at any point in the pumping procedure. In \cref{fig: Zak Phase and Gap Modes}(d), the novel feature of multilevel pumping can be observed, as was pointed out in Ref.~\cite{Yoshii2021TopologicalIndex}. The reason for this behavior can be understood from a real-space-renormalization approach. The instantaneous state at the largest gap can mapped to a state in a smaller gap, with renormalized Hamiltonian parameters. In Figs.~\ref{fig: Zak Phase and Gap Modes}(e) and \ref{fig: Zak Phase and Gap Modes}(f), similar behavior is observed for the Tribonacci chain, with multilevel charge pumping and edge state crossing points appearing around $t=T/4 $ and $t=T/4$. Finally, Figs.~\ref{fig: Zak Phase and Gap Modes}(g) and \ref{fig: Zak Phase and Gap Modes}(h) show the results for the Thue-Morse chain, also indicating multi-level charge pumping and the edge state crossing points for the largest gaps. Unlike the two quasiperiodic chains, the Thue-Morse chain is inversion symmetric. This is because the Thue-Morse word is a palindrome for even generations of the word. In \cref{Sec: Topo Signatures Aperiodic}, we will further investigate the palindromic realizations across all the models.

To ensure that these features are not dependent on the choice of unit cell, the same calculations were performed using different unit cell configurations. The results are presented in \cref{App: B}.

In order to corroborate the claims on multilevel pumping,  we also calculate the amount of charge pumped at any time, which is given by the partial sum of \cref{eq: Time Bott Index}, i.e.
\begin{equation*}
    Q(t) = \frac{1}{2\pi}\sum_{t'=0}^t \Im\Tr\log(\Tilde{V}_{t',t'+\Delta t}\Tilde{U}_{t'+\Delta t}\Tilde{V}^\dagger_{t',t'+\Delta t}\Tilde{U}^\dagger_{t'}).
\end{equation*}
The results are shown in \cref{Fig: Cumulative Charge All models}. In \cref{Fig: Cumulative Charge All models} (a), the periodic modulation is shown at half-filling, as there is only one gap. This is single-level pumping. In Figs.~\ref{Fig: Cumulative Charge All models} (b) - (d), the charge pumped at three different fillings for the Fibonacci, Tribonacci, and Thue-Morse modulations is shown, respectively. In each case, we see that the charge gradually increases to a maximum of 1 at the end of the pumping cycle for different fillings, a hallmark of quantized multilevel charge pumping. We also note that the form of the curve tends towards a step-like function for decreasing filling in all three cases, generalizing the observations made in Ref.~\cite{Yoshii2021TopologicalIndex}.  

For each case, we calculated the Berry phase for a filling corresponding to the largest gap below $E=0$. This gap is indicated by the green shaded area in \cref{fig: Zak Phase and Gap Modes}. In each case, we see that $\phi(t)=\pi$ exactly where the edge modes cross, indicating that the system could be in a nontrivial 1D topological phase (provided it possesses inversion symmetry), with an anomalous polarization of $P=1/2$. This motivates us to investigate the $\phi=\pi$ phase in more detail to show the anomalous topological response appearing at those points. 

\section{Topological Signatures in Aperiodic Systems} \label{Sec: Topo Signatures Aperiodic} 

Before investigating these topological signatures, we shall describe inversion-symmetric realizations of aperiodic chains in more detail. 

\subsection{Inversion-symmetric aperiodic chains}
As stated earlier, the Thue-Morse chain pumping cycle happens between two realizations of inversion-symmetric chains. This is because of the property that finite Thue-Morse words $W_n$ of even generation $n$ form perfect palindromes.  A snapshot of the Hamiltonian when the Berry phase is $\phi=\pi$ reveals that the Hamiltonian is of the form 
\begin{equation}\label{Eq: mirror symmetric Hamiltonian}
    H = \sum_{j}V_jc^\dagger_jc_j + \Tilde{t}\sum_{j}c^\dagger_{j}c_{j+1} + \text{h.c.},
\end{equation}
where $V_j = V_A, V_B$, distributed according to the Thue-Morse sequence, and $\Tilde{t}=2$ (arbitrary units). The on-site potentials take on the values $V_A = -0.25\Tilde{t}$ and $V_B = 0.5\Tilde{t}$. On the other hand, when $\phi=0$, $\Tilde{t}$ is still the same, but now the on-site potentials are $V_A = 0.25\Tilde{t}$ and $V_B = -0.5\Tilde{t}$. i.e., the sign has been switched. 

With this in mind, we proceed similarly for the Fibonacci and Tribonacci chains. However, we must select palindromic sections of the finite words. For the Fibonacci chain, one can prove by induction that for any generation $n$, the word $W_n = P_nxy$, where $P_n$ is a palindrome and $xy=AB$ or $xy=BA$, depending on the generation. Therefore, we shall work with Fibonacci chains of generation $n$ and omit the last two letters. 

For the Tribonacci chain, it is also known \cite{Tan2007SomeSequence} that one can factor the generation $n\geq 1$ word as $W_n = P_nE_n$, where 
\begin{equation*}
    P_n = W_{n-1}W_{n-2}\cdots W_1W_0
\end{equation*}
is a palindrome and $E_n$ is a word of length $|E_n| = (|W_n|-|W_{n-2}+3)/2$. Thus, we shall be working with the palindrome $P_n$ in this case as well. In all cases, we will use the simple hopping Hamiltonian \cref{Eq: mirror symmetric Hamiltonian}.

\begin{figure}
    \centering
    \includegraphics[width=\columnwidth]{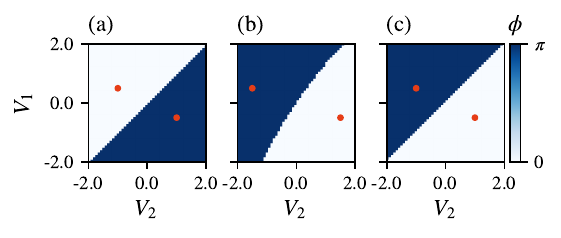}
    \caption{Phase diagram of the inversion-symmetric realizations of aperiodic chains. The generations chosen are (a) $n=9$ for the Fibonacci chain, (b) $n=8$ for the Tribonacci chain, and (c) $n=6$ for the Thue-Morse chain. The hopping parameter has been set to $\Tilde{t}=1$ for the Fibonacci and Tribonacci chains, while it has been set to $\Tilde{t}=2$ for the Thue-Morse chain to reflect the phase achieved in the pumping cycle shown in \cref{fig: Zak Phase and Gap Modes}(h).}
    \label{fig: phase diagrams Berry phase}
\end{figure}

In \cref{fig: phase diagrams Berry phase}(a), we show the phase diagram depicting the quantization of the Berry phase in the $(V_A, V_B) \equiv (V_1, V_2)$ plane for the Fibonacci modulation. Figures \ref{fig: phase diagrams Berry phase}(b) and (c) show the same for the Tribonacci and the Thue-Morse chains. Since the Tribonacci chain has three on-site potentials, we set the most recurrent one equal to $V_A = 0$ and the other two to $(V_B, V_C) = (V_1, V_2)$. \\

Given the considerations laid down at the end of \cref{Sec: TCPAS}, we expect a nontrivial insulating phase in 1D, with the corresponding protected anomalous boundary responses at exactly the points where the finite approximants show inversion symmetry. 
Nevertheless, it was shown that is possible to adiabatically transform a quasiperiodic system, such that the open chain has its edge states pushed into the bulk. Thereby, one may identify the phases as being topologically equivalent to the trivial insulator \cite{Madsen2013TopologicalStructures,Rontgen2019LocalChains}, which renders these topological phases very fragile. This is a common feature of 1D topological insulators, as the SSH chain without chiral symmetry presents very weakly protected edge modes. The presence of both chiral and inversion symmetries renders them more robust, as the former pins down the degenerate edge states at $E=0$. When chiral symmetry is lifted, one can adiabatically deform the Hamiltonian to move the edge states into bulk bands. 

In the following, we will show the existence of these delicate topological phases by using two typical signatures of nontrivial topology in 1D, which do not require the calculation of bulk topological invariants. The first one is the polarization response of an open system, and the second is the ES degeneracy. 

\subsection{Polarization}
\begin{figure}[!b]
    \centering
    \includegraphics[width=\columnwidth]{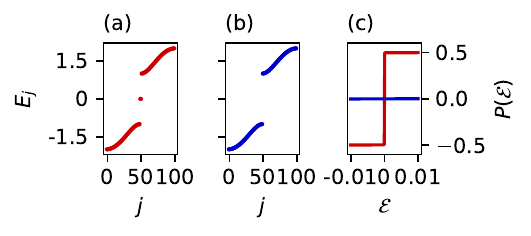}
    \caption{Energy and polarization of the SSH chain. (a) The OBC spectrum of the SSH chain in the topological phase: two degenerate edge states are pinned at the Fermi energy. (b) The OBC spectrum in the trivial phase: all states belong to the bulk. (c) The two different polarization responses, in the trivial phase (blue) and in the topological phase (red). The dimerization parameter in \cref{Eq: SSH chain} has been set to $\delta=\pm0.5$ [$\delta<0$ ($\delta>0$): topological (trivial) phase].}
    \label{fig: Anomalous SSH}
\end{figure}

\begin{figure}[!hbt]
    \centering
    \includegraphics[width=\columnwidth]{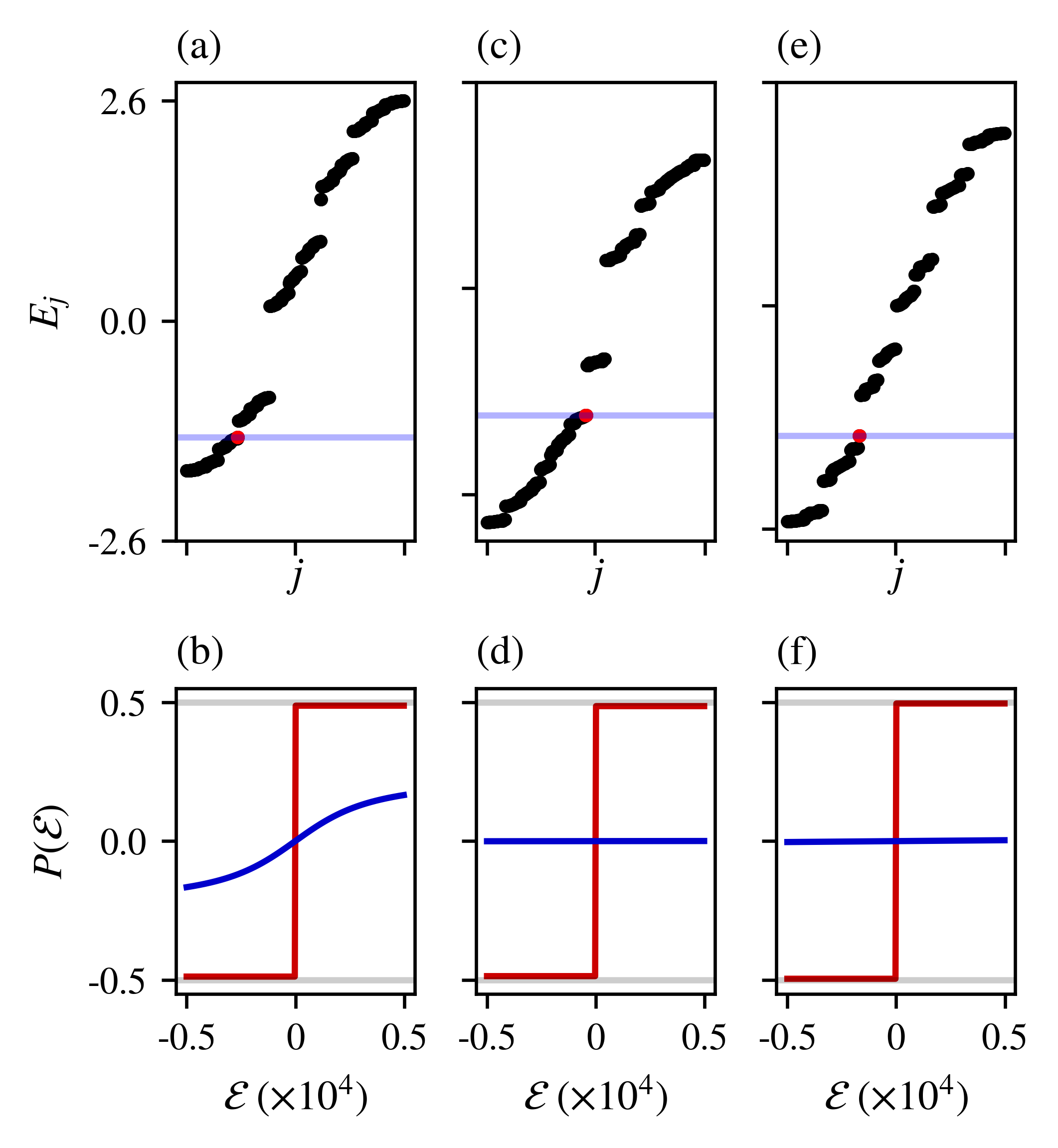}
    \caption{Polarization response of the aperiodic chains. OBC spectra (in the nontrivial phase) and polarization responses of the (a,b) Fibonacci chain, (c,d) Tribonacci chain and (e,f) Thue-Morse chain. The fillings are chosen such that the Fermi energies (light blue line) lie in the largest gap in each case. In all cases, the polarization exhibits an anomalous behavior around $\mathcal{E}=0$, with a jump from $-0.5$ to $0.5$ caused by the eigenstates colored in red in (a), (c), and (e) whenever the system is in the $\phi=\pi$ phase. On the other hand, the polarization does not show a sudden jump in the $\phi=0$ phase (in blue). The system sizes are $N = F_{14} - 2 = 608$, for the Fibonacci chain, $N=T_{11}-|E_{11}|=325$ for the Tribonacci chain, and $N=2^8=256$ for the Thue-Morse chain. The number of unit cells was fixed at 1, corresponding to a single aperiodic cell.}
    \label{fig: Aperiodic polarization}
\end{figure}
We use the SSH chain introduced in \cref{Eq: SSH chain} as a benchmark for anomalous polarization responses. We expect that as soon as an infinitesimal electric field is turned on, the polarization has a value of $|P|=0.5$ when the chain is in the topological phase. In order to probe the boundary response, we add an electric field along the chain, contributing $H_{\mathcal
{E}}=\mathcal
{E}\hat{X}$ to the Hamiltonian, where $\mathcal
{E}$ is the electric field strength, and the position operator is chosen to be defined as
\begin{equation}\label{Eq: gen position operator}
\begin{split}
    \begin{multlined}\hat{X}=
        \sum_{i=1}^{N_c}\sum_{j=1}^{N_s}\left(-\frac{2j-1}{2N_s}-\frac{N_c}{2}+i\right)c_{i,j}^\dagger c_{i,j},
    \end{multlined}
\end{split}
\end{equation}
where $N_c$ is the number of cells and $N_s$ is the number of sites in a cell. This is a generalization of the definition given in the case of the SSH chain in Ref.~\cite{Aihara2020AnomalousPhase}, which takes the form
\begin{equation*}
\begin{split}
    \begin{multlined}\hat{X}=
        \sum_{j=1}^N\left(-\frac{3}{4}-\frac{N}{2}+j\right)c_{2j-1}^\dagger c_{2j-1}\\ +\left(-\frac{1}{4}-\frac{N}{2}+j\right)c_{2j}^\dagger c_{2j},
    \end{multlined}
\end{split}
\end{equation*}
where $N$ is the number of unit cells. 

The position operator $\hat{X}$ is chosen such that $x=0$ lies in the middle of the chain. Note that the total amount of sites is $L=N_sN_c$. The full Hamiltonian is then $H=H_{\text{system}}+H_{\mathcal
{E}}$, where $H_{\text{system}}$ corresponds to any system that we wish to study. The dielectric response is given by the polarization \cite{Aihara2020AnomalousPhase}, 
\begin{equation}
    P(\mathcal
{E})=-\frac{1}{L-1}\sum_{n=1}^N\frac{\partial E_n}{\partial \mathcal
{E}}
\end{equation}
where $E_n\in\text{Spect}(H)$. In the limit $\mathcal
{E}\to0$, $|P|$ should agree with the polarization mentioned in \cref{sec:Intro}. In that limit, the adiabatic theorem applies, and one can choose a  temporal gauge such that the minimal coupling of the electric field allows for a temporal sweep of the complete BZ \cite{Bradlyn2022LectureTopology}. 

More generally, using a periodic approximation for the aperiodic systems, we shall make use of \cref{Eq: gen position operator}, where each cell represents an approximant of the aperiodic structure.

\begin{figure*}[!hbt]
    \centering
    \includegraphics[width=\textwidth]{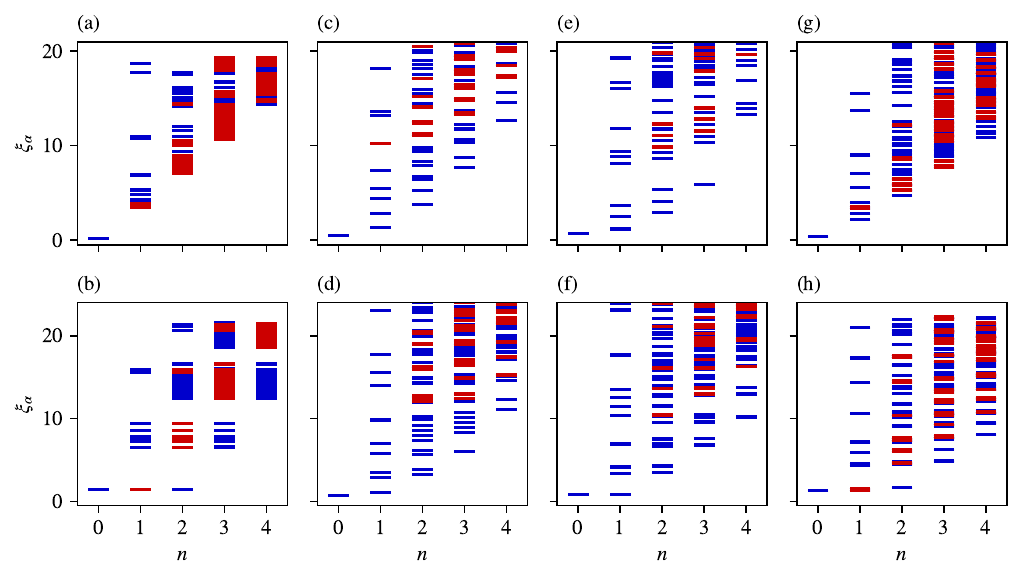}
    \caption{First few eigenvalues of the ES plotted against the total particle number of the eigenvalue configuration. The first row indicates all the trivial realizations of the inversion-symmetric chains, and the lower one indicates all the topological realizations. (a,b) The SSH chain. (c,d) the
    Fibonacci chain. (e,f) The Tribonacci chain. (g,h) The Thue-Morse chain. A red bar indicates
    at least double degeneracy at the given particle number. However, an eigenvalue $\xi_\alpha$ can also appear for a different particle number configuration, yielding another type of degeneracy (see \cref{App: A} for more details). Note that for the Fibonacci chain, the eigenvalues exhibit quasi-degeneracy due to finite-size effects. As the system size increases, this degeneracy becomes more pronounced.}
    \label{fig: Entanglement Spectrum Aperiodic}
\end{figure*}

\paragraph{SSH Chain.} The spectra and polarization of the SSH chain are shown in \cref{fig: Anomalous SSH}. The OBC spectra for the topological (red) and the trivial (blue) phase are shown in Figs.~\ref{fig: Anomalous SSH}(a) and \ref{fig: Anomalous SSH}(b), respectively.  The anomalous response of the SSH chain in the topological phase is shown in \cref{fig: Anomalous SSH}(c) in red, while the the response in the trivial phase is shown in blue. These results agree with the quantized Berry phase of $\phi=\pi,0$, respectively.

\paragraph{Aperiodic Chains.} We will now analyze the responses of three different aperiodic chains, each one being a representative of a different class of aperiodicity. 

The generic system Hamiltonian will be of the form \cref{Eq: mirror symmetric Hamiltonian}, where the on-site potentials are modulated, and the hoppings are constant. The choices for $(V_A, V_B)$ are shown in red in \cref{fig: phase diagrams Berry phase}, and the system sizes are also the same.

For the Fibonacci modulation, \cref{fig: Aperiodic polarization}(a) depicts the OBC energy levels, with a choice of Fermi energy indicated by the blue line. In \cref{fig: Aperiodic polarization}(b), the corresponding polarization $P(\mathcal{E})$ response is shown. There is a clear anomalous polarization, as indicated by the sudden jump from $P(\mathcal{E})=-0.5$ to $P(\mathcal{E})=0.5$ near $\mathcal {E}\approx0$, which is caused by the eigenstates colored in red, located at the Fermi level (light blue) in the $\phi=\pi$ phase. On the other hand, the polarization response does not show a sudden jump in the $\phi=0$ phase (in blue). The same behavior is observed for the Tribonacci modulation in Figs.~\ref{fig: Aperiodic polarization}(c) and \ref{fig: Aperiodic polarization}(d). For the Thue-Morse chain, in Figs.~\ref{fig: Aperiodic polarization}(e) and \ref{fig: Aperiodic polarization}(f), we also observe a similar behavior.
We note that the results plotted in \cref{fig: Aperiodic polarization} do not show a perfect equality $|P|=0.5$ (topological), or $P=0$ (trivial), which we suspect might be due to finite-size effects of the numerical implementation of the calculations. All system sizes have been chosen to be the same as those used to calculate the phase diagram in \cref{fig: phase diagrams Berry phase}, with parameter values as indicated by the red points in the figures. Namely, we have chosen (a) $(V_1, V_2) = (\mp0.5\Tilde{t}, \pm1\Tilde{t})$, (b) $(V_1, V_2) = (\pm0.5\Tilde{t}, \mp1.5\Tilde{t})$, and (c) $(V_1, V_2) = (\mp0.25\Tilde{t}, \pm0.5\Tilde{t})$ (in arbitrary units) for the topological and trivial phases, respectively.

\subsection{Entanglement Spectrum}

The second signature of nontrivial topological states that we will use is the degeneracy structure of the ES. There exists a one-to-one correspondence between the topological boundary modes of an open system and the degeneracy of all eigenvalues of the ES of a subsystem taken deep in the bulk of an extended system \cite{Fidkowski2010EntanglementSuperconductors}. In the following section, we briefly recall how to calculate the ES and the method we used to obtain the results. We then show that the signatures of nontrivial topology are present in both the SSH chain and the aperiodic chains.

Let us consider the ground state of a fermionic insulator, described by a quadratic Hamiltonian with a gapped single-particle spectrum. The many-body ground state of such a system is a pure product state, which can be written as 
\begin{equation*}
    \ket{\Psi_0}=\prod_{n<n_F}\alpha^\dagger_n\ket{0},
\end{equation*}
where the operator $\alpha^\dagger_n$ creates a particle in the $n^\text{th}$ eigenstate of the Hamiltonian. 
The density matrix for this pure state is simply given by $\rho_0=\ket{\Psi_0}\bra{\Psi_0}$. Given a certain bi-partition of the chain, we want to probe the entanglement between two subsystems. Let the two systems be labeled by $K$ and $L$. Then, the ES is defined to be the set of the negative logarithm of the eigenvalues of the reduced density matrix $\rho_L$ (or equivalently $\rho_K$), i.e., 
\begin{equation}
    \text{ES}=\{\xi_\alpha\in \text{Spect}(-\log\rho_L)\ | \ \rho_L=\Tr_K\rho\},
\end{equation}
where $\Tr_K$ is a partial trace over the subsystem $K$. In the following, we will plot the ES against particle number configuration (see \cref{App: A} for more details on how to calculate the ES).

The ES of the SSH chain is shown in Figs.~\ref{fig: Entanglement Spectrum Aperiodic}(a) and \ref{fig: Entanglement Spectrum Aperiodic}(b). The difference between the trivial \cref{fig: Entanglement Spectrum Aperiodic}(a) and topological \cref{fig: Entanglement Spectrum Aperiodic}(b) phases is in the increased degeneracy of all eigenvalues in the topological phase. Restricting ourselves to the lowest eigenvalue, we see that its degeneracy goes from $D=1$ to $D=4$ (the red bar indicates that the degeneracy is at least of order 2). These results are already known from Ref. \cite{Sirker2014BoundaryModel}. We shall use the degeneracy of the lowest eigenvalue as an indicator of nontrivial topological behavior at the boundary of our 1D models.

In Figs.~\ref{fig: Entanglement Spectrum Aperiodic} (c)-(h), the ES of the aperiodic chains are shown. For each case, at least a double degeneracy of the lowest eigenvalue can be observed in the bottom row of the figure.
It was already known that the bulk entanglement entropy of the Fibonacci quasicrystal carries some type of signatures of the gap labels \cite{Rai2021BulkQuasicrystal}. Our findings provide an indication of the nontrivial topology that could arise at the boundary when the in-gap modes are inversion-symmetric partners. 
    
\section{Conclusion} \label{Sec: Conclusion}
The study of topological phases in quasicrystals has attracted a significant amount of attention in the last decade, and there have been many interesting experimental and theoretical observations \cite{Kraus2012TopologicalModel,Kraus2012TopologicalQuasicrystals,Verbin2013ObservationQuasicrystals,Verbin2015TopologicalQuasicrystal,Kellendonk2019BulkBoundaryModels}. At the same time, it is known that topological phases in 1D band insulators do not exist, except when inversion symmetry gives rise to an obstructed atomic phase. The latter results in a quantized polarization response due to fractionalized charges at the boundary, resulting from a filling anomaly \cite{Bradlyn2017TopologicalChemistry}. 

In this work, we put the recent observation of topological states in quasicrystals in context and show that multilevel charge pumping is a generic feature of aperiodic chains rather than being specific to quasiperiodic models. When these models admit inversion symmetry, such as the Thue-Morse chain, the pumping process cycles through two topologically distinct phases of the inversion-symmetric configurations. For many chains that admit palindromic factors, it is natural to expect nontrivial topological phases for finite approximants of the aperiodic systems. By calculating three typical signatures of topology in aperiodic chains, we have found truly 1D topological phases characterized by a quantized Berry phase for the periodic approximants. In addition, we have shown that the Fibonacci and Tribonacci quasicrystals and the Thue-Morse chain -- representing different classes of aperiodic systems, exhibit anomalous polarization responses and that their ES possess topological eigenvalue degeneracy. 

Just like for their periodic counterparts, topological charge pumping is very robust as it does not necessitate the presence of any symmetry. The spectral flow of the edge modes across bulk gaps is a topologically protected phenomenon resulting from the bulk-boundary correspondence. If the bulk gap does not close, any perturbations can be added to the system, and this spectral flow will remain intact. 
However, when constrained by additional inversion symmetry, individual realizations of these chains exhibit significant sensitivity to both disorder and open boundary conditions. Once again, like for their periodic counterparts, the degenerate edge modes can be easily disrupted by disorder or edge perturbations. Their robustness is confined to the model's parameter space, where the Berry phase satisfies $\phi=\pi$. For additional robustness, protection from a spectral symmetry—such as the chiral symmetry in the SSH chain—is required. However, such symmetries do not provide protection in any of the models considered in this work.

Our results do not hold for the Rudin-Shapriro modulation because the pumping protocol does not perform an adiabatic evolution of the system. Moreover, it is known that this sequence does not admit any palindromic factor \cite{Allouche1997SchrodingerPalindromic}. This might be due to the different topology of the energy and Fourier spectra. As a possible outlook, it would be interesting to understand the relationship between the energy and Fourier spectra, the generic property of multilevel pumping, and the existence of inversion-symmetric aperiodic insulators in general.
	
\begin{acknowledgments}
    We thank K. Dajani, L. Eek and Z. Osseweijer for fruitful discussions.
    This publication is part of the project TOPCORE with project number OCENW.GROOT.2019.048 which is financed by the Dutch Research Council (NWO).
\end{acknowledgments}

 \bibliography{Refs}

\appendix

\begin{widetext}

\section{Berry phase and edge state level crossing}\label{App: B}

As stated earlier in the main text, the time at which the level-crossing happens depends on the unit-cell choice. However, as we will empirically show in this appendix, the level crossing seems to always happen when the time-dependent Berry phase is equal to $\pi$. 
Before we do that, let us briefly explain how the unit cell is chosen for our calculations. As an example, we show the procedure for the Fibonacci word of generation $N=9$, with $F_9=55$ letters. Four choices of unit cells will be taken, where each successive one is shifted by a quarter word length (floored to the nearest integer), as shown below
\begin{equation*}
\begin{split}
    \text{Word: }& ABAABABAABAAB\underset{{\uparrow}}{\ } ABAABABAABAABA\underset{{\uparrow}}{\ }BAABAABABAABAB\underset{{\uparrow}}{\ }AABAABABAABABA, \\
    \text{Cell 1: }& ABAABABAABAABABAABABAABAABABAABAABABAABABAABAABABAABABA, \\
    \text{Cell 2: }& ABAABABAABAABABAABAABABAABABAABAABABAABABAABAABABAABAAB, \\
    \text{Cell 3: }& BAABAABABAABABAABAABABAABABAABAABABAABAABABAABABAABAABA, \\
    \text{Cell 4: }& AABAABABAABABAABAABABAABAABABAABABAABAABABAABAABABAABAB.
\end{split}
\end{equation*}
The first unit cell starts from the beginning of the word until its end. The second unit cell starts from where the first arrow points and winds around the word back to the last letter before the arrow. The same is done for the third unit cell, starting from the second arrow and so on... This procedure will be applied to each aperiodic word in this work. 

Since the systems under consideration are multiband insulators with valence bands that can cross each other, we must employ a more general definition for the Berry phase that we calculated numerically. To this end, we use \cite{Resta1994MacroscopicApproach}
\begin{equation*}
    \phi = -\text{Im}\log\prod_{k}\sum_{n=1}^{m}\ket{\psi_n(k)}\bra{\psi_n(k+\delta k)},
\end{equation*}
where $m$ is the number of filled bands and $\ket{\psi_n(k)}$ is the Bloch state of the $n^\text{th}$ band, with the following periodic boundary condition imposed in $k$-space,
\begin{equation*}
    \ket{\psi_n(k_N)}_{j} = \exp\left(-2\pi i x_j\right)\ket{\psi_n(k_0)}_{j}.
\end{equation*}
Here, the notation refers to the $j^\text{th}$ component of the discrete Bloch state, and $x_j$ is the position of the $j\text{th}$ lattice point. 

\begin{figure*}[!t]
    \centering
    \includegraphics{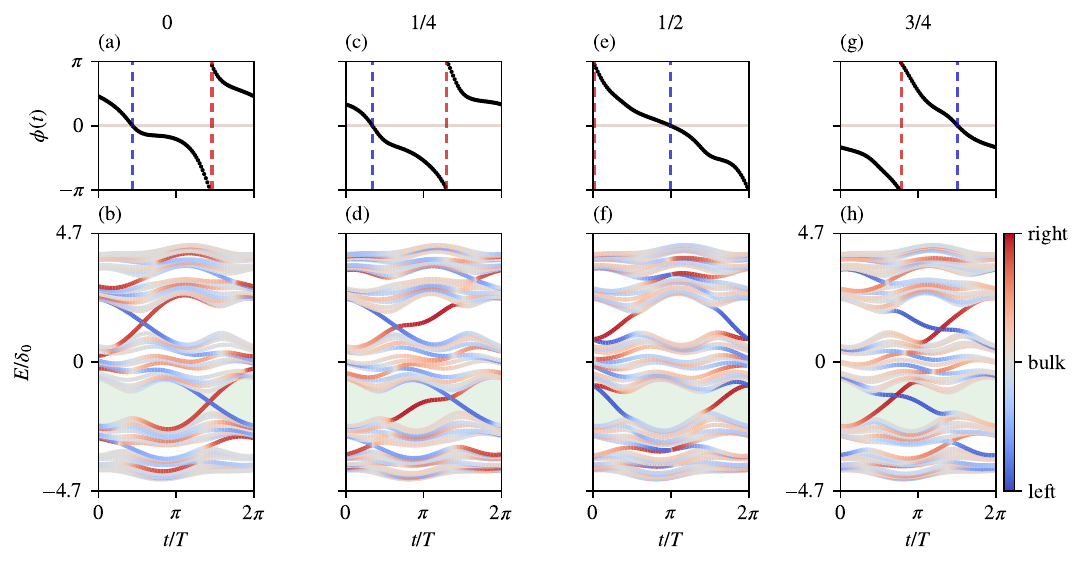}
    \caption{Berry Phase and level crossing for the Fibonacci chain. Each column corresponds to a different choice of a unit cell, where the number above each column represents the fraction of the length of the Fibonacci word used as a starting point to generate the unit cell. The green shaded area in the bottom row represents the chosen bulk gap for the Berry phase calculation. The system size is $N=55$. The red (blue) dashed lines mark the times when $\phi=\pi$ ($\phi=0$).}
    \label{fig: Fibo Berry Crossing}
\end{figure*} 
The results are plotted in Figs.~\ref{fig: Fibo Berry Crossing}, \ref{fig: Tribo Berry Crossing}, and \ref{fig: Thue-Morse Berry Crossing}. In each case, we see in the green shaded region, corresponding to the bulk gap considered, that the edge states cross exactly when $\phi=\pi$, as indicated by the red dotted lines in the upper row. 

\begin{figure*}[!hbt]
    \centering
    \includegraphics{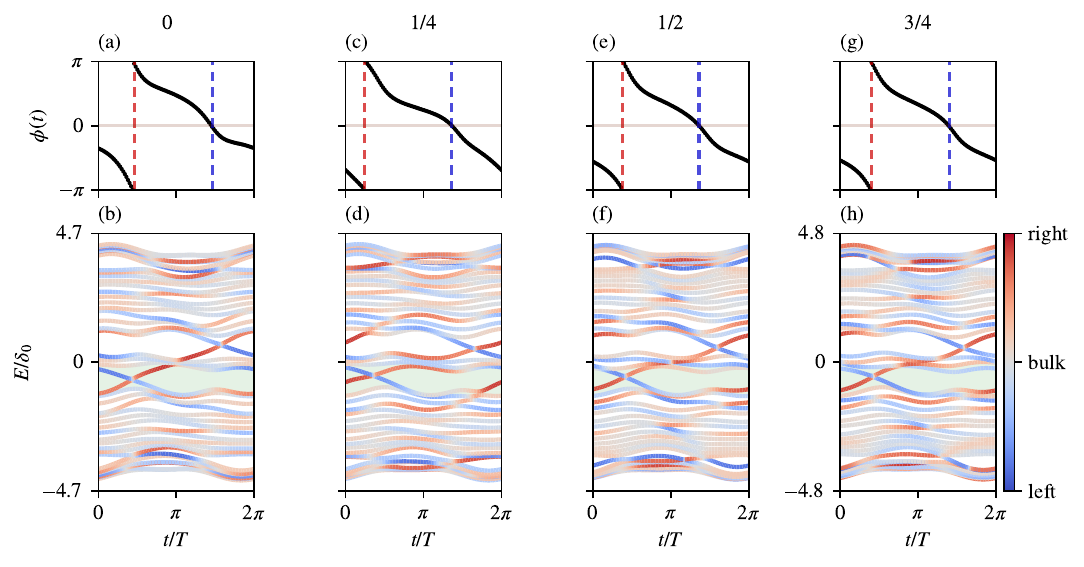}
    \caption{Berry Phase and level crossing for the Tribonacci chain. The system size is $N=44$.}
    \label{fig: Tribo Berry Crossing}
\end{figure*}

\begin{figure*}[!hbt]
    \centering
    \includegraphics{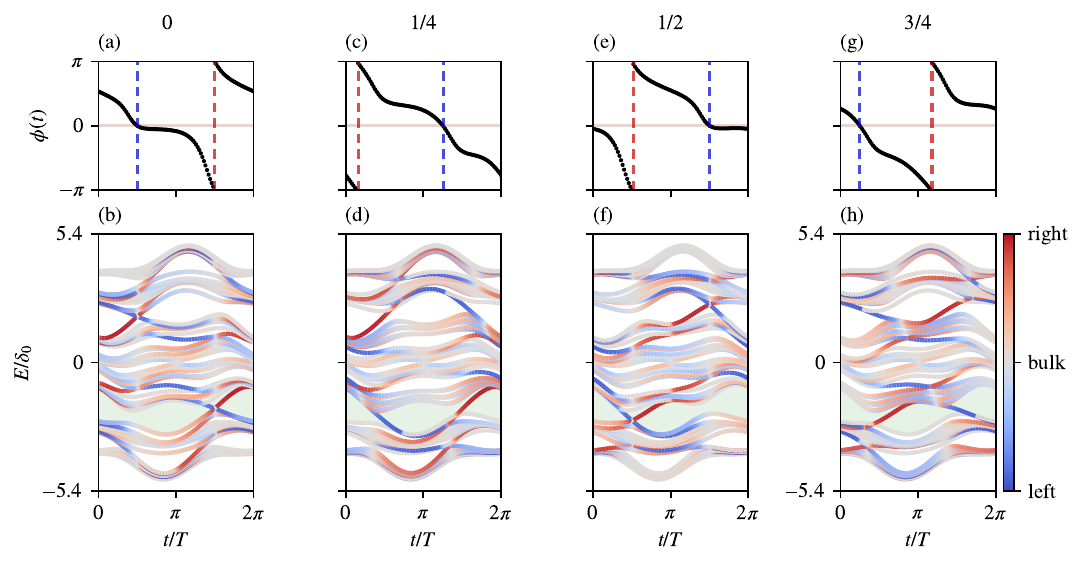}
    \caption{Berry Phase and level crossing for the Thue-Morse chain. The system size is $N=32$.}
    \label{fig: Thue-Morse Berry Crossing}
\end{figure*} 

\clearpage

\section{Entanglement Spectrum of Quadratic Hamiltonians}\label{App: A}
For quadratic Hamiltonians, it is known that the reduced density matrix takes the exponential form \cite{Cheong2004Many-bodyFermions}
\begin{equation*}
    \rho_L=\det\mathcal{M}_L \exp\left[\sum_{i,j=1}^{|L|}\log\left(G_L\mathcal{M}_L^{-1}\right)_{ij}c_i^\dagger c_j\right],
\end{equation*}
where $\mathcal{M}_L=\mathbbm{1}-G_L$ and $(G_L)_{ij}=\langle c^\dagger_ic_j\rangle$ is the correlation function restricted to subsystem $L$, of size $|L|$. Here, $c^\dagger_i$ creates an electron at site $x_i\in L$.
Diagonalizing $G_L=S\ \text{diag}(\lambda_1,\cdots,\lambda_n) \ S^{\dagger}$, we obtain 
\begin{equation*}
    \rho_L=\exp\left[\sum_{n=1}^{|L|}\log(1-\lambda_n)+\sum_{n=1}^{|L|}\log\left(\frac{\lambda_n}{1-\lambda_n}\right)d_n^\dagger d_n\right],
\end{equation*}
where $d_n=\sum_{l}S_{nl}c_l$, $\lambda_n$'s are eigenvalues of the correlation matrix $G_L$. 
The density matrix is an operator acting on states in Fock space
\begin{equation*}
    \mathcal{F}=\mathcal{H}_0\oplus\mathcal{H}_1\oplus\mathcal{H}_2\oplus\cdots\oplus\mathcal{H}_{|L|},
\end{equation*}
where each $\mathcal{H}_n=\mathcal{H}_1^{\otimes n}$ is a tensor product of single-particle Hilbert spaces. In this case, the single-particle Hilbert space is spanned by the $|L|$ eigenstates of the correlation matrix $G_L$. Since we have fermionic particles, the occupation of each single-particle state is either $0$ or $1$. Let us label an occupation configuration by $n^{(\alpha)} = \{n_1^{(\alpha)},n_2^{(\alpha)},\cdots,n_{|L|}^{(\alpha)} \}$, where each $n_j^{(\alpha)}=0$ or 1, and the ordering can be taken in terms of increasing $\lambda_n$. Note that there are $2^{|L|}$ such configurations, giving the size of the Fock space $\mathcal{F}$. The density matrix can be written in the basis of Fock-state vectors with these ``good'' quantum numbers (good with respect to $G_L$) as
\begin{equation*}
    \rho_L = \sum_{\alpha=1}^{2^{|L|}}\xi_\alpha\ket{n_1^{(\alpha)},n_2^{(\alpha)},\cdots,n_{|L|}^{(\alpha)}}\bra{n_1^{(\alpha)},n_2^{(\alpha)},\cdots,n_{|L|}^{(\alpha)}},
\end{equation*}
where the entanglement eigenvalues are given by 
\begin{equation*}
    \begin{split}
           \xi_\alpha=-\sum_{j=1}^{|L|}\log(1-\lambda_j)\left[\frac{1+(-1)^{n_j^{(\alpha)}}}{2}\right] - \sum_{j=1}^{|L|}\log(\lambda_j)\left[\frac{1-(-1)^{n_j^{(\alpha)}}}{2}\right].
    \end{split}
\end{equation*}
To each configuration $n^{(\alpha)}$, one can associate a total number of particles in that configuration, which we label $n=\sum_{j=1}^{|L|}n_j^{(\alpha)}$. 
Naturally, it is possible for different configurations $n^{(\alpha)}$ to yield the same number of particles. It is also possible for two different configurations to yield exactly the same entanglement eigenvalue $\xi_\alpha$, constituting a degeneracy in the density matrix. In order to properly visualize the entanglement spectrum in \cref{fig: Entanglement Spectrum Aperiodic}, we plot it against the particle number $n$.
The entanglement spectrum degeneracy can thus occur in two different ways: 1) with configurations yielding the same total number of particles $n$ [which is colored in red in fig.(7)] and 2) with configurations yielding a different number of particles $n$. 

\end{widetext}

\end{document}